\documentclass[pdflatex,sn-mathphys]{sn-jnl}
\usepackage{color}
\usepackage[normalem]{ulem}
\usepackage{cleveref}
\usepackage{lineno}

\jyear{2021}

\unnumbered

\begin{document}

\title[Nature Energy (Analysis Type)]{Abominable greenhouse gas bookkeeping casts serious doubts on climate intentions of oil and gas companies}

\author*[1]{\fnm{Sergio} \sur{Garcia-Vega}}\email{sergio.garcia-vega@ucd.ie}
\equalcont{These authors contributed equally to this work and are listed alphabetically.}

\author[1,2]{\fnm{Andreas G. F.} \sur{Hoepner}}\email{andreas.hoepner@ucd.ie}
\equalcont{These authors contributed equally to this work and are listed alphabetically.}

\author[3,4]{\fnm{Joeri} \sur{Rogelj}}\email{j.rogelj@imperial.ac.uk}
\equalcont{These authors contributed equally to this work and are listed alphabetically.}

\author[5]{\fnm{Frank} \sur{Schiemann}}\email{frank.schiemann@uni-bamberg.de}
\equalcont{These authors contributed equally to this work and are listed alphabetically.}

\affil*[1]{\orgdiv{Smurfit Graduate Business School}, \orgname{University College Dublin}, \orgaddress{\country{Republic of Ireland}}}

\affil[2]{\orgdiv{Independent Member}, \orgname{Platform for Sustainable Finance, DG FISMA}, \orgaddress{\country{EU}}}

\affil[3]{\orgdiv{Centre for Environmental Policy and Grantham Institute for Climate Change and the Environment}, \orgname{Imperial College London}, \orgaddress{\country{UK}}}

\affil[4]{\orgdiv{Energy, Climate and Environment Program}, \orgname{International Institute for Applied Systems Analysis}, \orgaddress{\country{Austria}}}

\affil[5]{\orgdiv{Faculty of Social Sciences, Economics, and Business Administration}, \orgname{University of Bamberg}, \orgaddress{\country{Germany}}}

\abstract{The Paris Agreement aims to reach net zero greenhouse gas (GHG) emissions in the second half of the 21st century, and the Oil \& Gas sector plays a key role in achieving this transition. Understanding progress in emission reductions in the private sector relies on the disclosure of corporate climate-related data, and the Carbon Disclosure Project (CDP) is considered a leader in this area. Companies report voluntarily to CDP, providing total emissions and breakdowns into categories. How reliable are these accounts? Here, we show that their reliability is likely very poor. A significant proportion of Oil \& Gas companies’ emission reports between 2010 and 2019 fail a `simple summation' mathematical test that identifies if the breakdowns add up to the totals. Companies' reports reflect unbalanced internal bookkeeping in 38.9\% of cases, which suggests worryingly low quality standards for data guiding the private sector’s contribution to halting climate change.}

\maketitle

Halting global warming and meeting the Paris Agreement requires global carbon dioxide ($CO_{2}$) emissions from human activities to be reduced to net zero and potentially even beyond~\cite{IPCC_AR6_WGI_2021}. The Oil \& Gas (O\&G) industry produced 56\% of all energy-related $CO_{2}$ in 2019, showing a need to curb its operational emissions in order to meet international climate goals~\cite{bouckaert2019world}. In general, the performance of O\&G companies on climate change is measured using greenhouse gas (GHG) emissions, which are widely used to track companies’ carbon footprint~\cite{dietz2021ambitious}. However, climate data is not without accuracy challenges~\cite{depoers2016voluntary,mateo2021influence}, which the United Nations attributes to the application of different reporting formats and inconsistency in the scope and timeliness of reporting~\cite{washingtonpost_climate}. 

Data quality can be measured in various ways. Researchers often use \textit{i)} measures of self-constructed indices, or \textit{ii)} disclosure ratings measures~\cite{lemma2019corporate}. CDP (formerly Carbon Disclosure Project) considers itself “the gold standard of environmental reporting with the richest and most comprehensive dataset on corporate\dots action" and certain stakeholders indeed credit it as the leader in terms of usefulness and quality for both investors and experts (\href{https://www.sustainability.com/thinking/rate-the-raters-2020/}{https://www.sustainability.com}). The CDP  disclosure score evaluates the comprehensiveness of companies' response to the CDP questionnaire, considering that some questions are conditional on others (e.g., firms are first asked if they can provide a breakdown of their emissions in more granular categories, whereafter they are asked to provide details for the identified breakdowns). 

The CDP scores are based on companies' answers to the CDP questionnaire and as such are based on self-reported information. One important element of CDP's questionnaire is the reporting of GHG emissions and the GHG emissions breakdowns for emissions directly resulting from a company's activities (\textit{Scope~1}), including their purchased energy use (\textit{Scope~2}), or additionally including emissions caused by the broader supply chain, distribution, and use of a company's products (\textit{Scope~3}). Although previous studies show that companies' response to CDP is increasing, the quality and content of disclosure are still insufficient to satisfy the requirements of investors~\cite{dragomir2012disclosure,peng2015corporate,li2019carbon}. That is, the reliability and comparability of GHG emission data is questioned by investors, which leads to the criticism that CDP disclosure contributes more to “green washing” than to improving corporate transparency~\cite{dragomir2012disclosure}. It is important to note, that despite such criticism, CDP is seen as the leading provider of corporate climate-related information.

Literature provides evidence that ratings of companies' Environmental, Social \& Governance (ESG) performance are not always converging, meaning that different rating agencies come to different conclusions for the same company \cite{chatterji2016ratings}. It seems, these difference can be accounted to elements of scope, weight across indicators, and measurement. This leads to the recommendation that raw data should be used, when feasible~\cite{berg2022aggregate}. The CDP has the advantage that it does not only provide aggregate scores, but the underlying raw data are also available from companies' answers. Therefore, investors are able to utilize the CDP dataset for detailled analyses based on the specific information contained in the questionnaire. Such analyses are especially relevant for high-emitting companies, such as in the O\&G sector. A key question, however, is whether the discolsed data is reliable and of decent quality. Therefore, we analysed the quality characteristics of direct GHG emissions that are owned or controlled by 33 O\&G companies (also known as \textit{Scope~1} emissions) in the CDP database between 2010 and 2019. 


\subsection{Basic Bookkeeping Test}
The accuracy of reported emissions is checked with simple summation tests that aim to identify the ability and intentions of the companies when voluntarily reporting their emissions. Our tests exploit the fact that, when reporting their \textit{Scope~1} emissions to the CDP, companies are encouraged to voluntarily report their total GHG emissions broken down into Activities, Business Units, Facilities, GHG Types, and Regions. 

The primary purpose of this mathematical test is to evaluate the ability of the analyzed companies in voluntarily disclosing their GHG emissions to the CDP (see Methods). The study analyzed the overall performance of these companies over the period of 2010 to 2019, using five breakdowns in Activity, Business, Facility, GHG, and Region. Our findings indicate that there is a significant proportion of reports that exhibit disparities between the reported global \textit{Scope~1} emissions and the total sum of \textit{Scope~1} emissions by breakdown. The upper section of \Cref{tab:test1_cdp_scope1_oilgas} focuses on the number of unique reports that demonstrate these discrepancies, which are categorized based on breakdowns (rows) and reporting years (columns). The lower section of \Cref{tab:test1_cdp_scope1_oilgas} presents the discrepancies in terms of the number of unique companies rather than the number of unique reports.

The last column in \Cref{tab:test1_cdp_scope1_oilgas} shows the average number of reports containing discrepancies during the examined 10-year period. In particular, the analyzed companies typically refrained from divulging their total gross global \textit{Scope~1} emissions categorized by Facility, i.e., only 5.6 companies disclosed their Facility breakdown every year. Moreover, a distressing 2.1 of those 5.6 firms (around $37\%$) are anticipated to display inconsistencies between the total sum of their Facility breakdown and their declared global emissions. This implies that only a small number of firms reported their emissions by Facility, and those that did frequently made errors in their reports. Conversely, most of the companies under analysis revealed their total gross global \textit{Scope~1} emissions by Region; that is, $21$ out of the $33$ O\&G companies disclosed their Region breakdown on average every year. This particular breakdown also had a low number of reporting discrepancies per year, averaging just $3.2$, making it the breakdown with the fewest mismatches (around $15\%$) among the five analyzed. In general, we find GHG emissions to represent unbalanced internal bookkeeping in $38.9\%$ of the cases, which was the worst in 2010 ($46.2\%$), while the year with the least unbalanced internal bookkeeping was 2012 ($25.9\%$). 

The performances of the analyzed companies in terms of emissions reporting to CDP are presented in detail in \Cref{tab:test1_scope1}. This evaluation compares the number of discrepancies identified between the companies' reported global \textit{Scope~1} emissions and the total sum of their emissions breakdowns. This provides a more in-depth view of the performance data presented in the previous analysis (\Cref{tab:test1_cdp_scope1_oilgas}), allowing a more granular assessment of individual companies. Note that the denominator values in this analysis do not exceed 10, as this represents the 10-year period between 2010 and 2019. For instance, Petrobras disclosed its emissions breakdown annually over the decade (denominator value of 10 in the Business, GHG, and Region columns). The numerator, on the other hand, indicates the number of erroneous reports submitted by each company within the same time frame. Thus, the higher the percentage, the worst the performance in terms of mismatched reports. In particular, Royal Dutch Shell had the highest number of erroneous reports (27 out of 29) during the 10-year period under review. Notably, the company did not disclose its emissions in the Activity and Facility breakdowns throughout the entire period from 2010 to 2019. However, the company did report its emissions every year (except for the Business breakdown) in other categories such as GHG and Region. Nevertheless, these reports were frequently found to be erroneous. In contrast, companies like Galp Energia SA and Total show the best performance as they did not have a single mismatched report on the 24 times that they reported in the same 10-year period.

The poor performance of companies such as Royal Dutch Shell, PTT, and Woodside Petroleum, where the percentage of discrepancies in their emissions reports is above $70\%$, is a matter of concern. It raises questions about their ability to use appropriate software or even basic tools like Excel spreadsheets to report their breakdowns accurately, about their management priorities and the quality standards applied to their internal management processes. Furthermore, their unreliable emissions reports pose a significant challenge for policymakers and the scientific community in designing effective policies to achieve net-zero GHG emissions. In essence, while disclosing emissions is crucial, it is equally crucial to do so accurately and properly. Failing to do so may result in incorrect policy decisions or false claims of progress towards reaching the net-zero goal.

\begin{sidewaystable}
\sidewaystablefn%
\begin{center}
\begin{minipage}{\textheight}
	{\footnotesize
		\caption{Overall Oil \& Gas companies performance (Test 1). Note, all reported values (mismatches, total reports, and total companies) have been obtained using the preprocessing procedure described in the methods  section.}
		\begin{tabular}{llcccccccccclcc}
			\hline
			\multicolumn{1}{c}{\multirow{2}{*}{\textbf{Breakdown}}} & \multicolumn{1}{c}{\multirow{2}{*}{\textbf{Parameter}}} & \multicolumn{10}{c}{\textbf{CDP Report}}                                                                                                                                                                                            &  & \multirow{2}{*}{\textbf{Total}} & \multirow{2}{*}{\textbf{Average}} \\ \cline{3-12}
			\multicolumn{1}{c}{}                                    & \multicolumn{1}{c}{}                                    & \textbf{2010}        & \textbf{2011}        & \textbf{2012}        & \textbf{2013}        & \textbf{2014}        & \textbf{2015}        & \textbf{2016}        & \textbf{2017}        & \textbf{2018}        & \textbf{2019}        &  &                                 &                                   \\ \hline
			\multirow{3}{*}{\textbf{Activity}}                      & Mismatch                                                &                      & 1                    & 4                    & 2                    & 3                    & 3                    & 3                    & 1                    & 1                    & 1                    &  & 19                              & 2.1                               \\
			& Percentage Mismatch                               &                      & 16.7                 & 40.0                 & 20.0                 & 30.0                 & 27.3                 & 30.0                 & 9.1                  & 33.3                 & 33.3                 &  & 25.7                            & 26.6                              \\
			& Total Reports                            &                      & 6                    & 10                   & 10                   & 10                   & 11                   & 10                   & 11                   & 3                    & 3                    &  & 74                              & 8.2                               \\
			&                                                         &                      &                      &                      &                      &                      &                      &                      &                      &                      &                      &  &                                 &                                   \\
			\multirow{3}{*}{\textbf{Business}}                      & Mismatch                                                & 5                    & 5                    & 2                    & 4                    & 4                    & 6                    & 6                    & 6                    & 2                    & 3                    &  & 43                              & 4.3                               \\
			& Percentage Mismatch                               & 27.8                 & 25.0                 & 10.5                 & 22.2                 & 18.2                 & 27.3                 & 27.3                 & 28.6                 & 20.0                 & 27.3                 &  & 23.5                            & 23.4                              \\
			& Total Reports                            & 18                   & 20                   & 19                   & 18                   & 22                   & 22                   & 22                   & 21                   & 10                   & 11                   &  & 183                             & 18.3                              \\
			&                                                         &                      &                      &                      &                      &                      &                      &                      &                      &                      &                      &  &                                 &                                   \\
			\multirow{3}{*}{\textbf{Facility}}                      & Mismatch                                                & 5                    & 2                    & 2                    & 2                    & 2                    & 2                    & 2                    & 1                    & 2                    & 1                    &  & 21                              & 2.1                               \\
			& Percentage Mismatch                               & 62.5                 & 28.6                 & 50.0                 & 50.0                 & 40.0                 & 40.0                 & 33.3                 & 20.0                 & 33.3                 & 16.7                 &  & 37.5                            & 37.4                              \\
			& Total Reports                            & 8                    & 7                    & 4                    & 4                    & 5                    & 5                    & 6                    & 5                    & 6                    & 6                    &  & 56                              & 5.6                               \\
			&                                                         &                      &                      &                      &                      &                      &                      &                      &                      &                      &                      &  &                                 &                                   \\
			\multirow{3}{*}{\textbf{GHG}}                           & Mismatch                                                & 7                    & 6                    & 4                    & 6                    & 6                    & 6                    & 7                    & 6                    & 6                    & 6                    &  & 60                              & 6.0                               \\
			& Percentage Mismatch                               & 31.8                 & 26.1                 & 18.2                 & 28.6                 & 27.3                 & 28.6                 & 33.3                 & 30.0                 & 33.3                 & 33.3                 &  & 28.8                            & 29.1                              \\
			& Total Reports                            & 22                   & 23                   & 22                   & 21                   & 22                   & 21                   & 21                   & 20                   & 18                   & 18                   &  & 208                             & 20.8                              \\
			&                                                         &                      &                      &                      &                      &                      &                      &                      &                      &                      &                      &  &                                 &                                   \\
			\multirow{3}{*}{\textbf{Region}}                        & Mismatch                                                & 3                    & 2                    & 1                    & 3                    & 3                    & 6                    & 3                    & 4                    & 4                    & 3                    &  & 32                              & 3.2                               \\
			& Percentage Mismatch                               & 14.3                 & 9.1                  & 4.3                  & 14.3                 & 13.6                 & 27.3                 & 14.3                 & 18.2                 & 22.2                 & 16.7                 &  & 15.2                            & 15.4                              \\
			& Total Reports                            & 21                   & 22                   & 23                   & 21                   & 22                   & 22                   & 21                   & 22                   & 18                   & 18                   &  & 210                             & 21.0                              \\
			&                                                         &                      &                      &                      &                      &                      &                      &                      &                      &                      &                      &  &                                 &                                   \\
			\multicolumn{2}{c}{\textbf{Total Mismatches}}                                                                     & 20                   & 16                   & 13                   & 17                   & 18                   & 23                   & 21                   & 18                   & 15                   & 14                   &  & \textbf{175}                    &                                   \\
			\multicolumn{2}{c}{\textbf{Total Reports}}                                                                        & 69                   & 78                   & 78                   & 74                   & 81                   & 81                   & 80                   & 79                   & 55                   & 56                   &  & \textbf{731}                    &                                   \\
			\multicolumn{2}{c}{\textbf{Percentage Mismatch}}                                                                           & 29.0                 & 20.5                 & 16.7                 & 23.0                 & 22.2                 & 28.4                 & 26.3                 & 22.8                 & 27.3                 & 25                   &  & \textbf{23.9}                   &                                   \\
			&                                                         & \multicolumn{1}{l}{} & \multicolumn{1}{l}{} & \multicolumn{1}{l}{} & \multicolumn{1}{l}{} & \multicolumn{1}{l}{} & \multicolumn{1}{l}{} & \multicolumn{1}{l}{} & \multicolumn{1}{l}{} & \multicolumn{1}{l}{} & \multicolumn{1}{l}{} &  & \multicolumn{1}{l}{}            & \multicolumn{1}{l}{}              \\ \hline
			\multicolumn{2}{l}{Total Companies}                                                            & \textbf{26}          & \textbf{27}          & \textbf{27}          & \textbf{25}          & \textbf{26}          & \textbf{27}          & \textbf{27}          & \textbf{27}          & \textbf{18}          & \textbf{18}          &  & 248                             & 24.8                              \\
			\multicolumn{2}{l}{Mismatch Companies}                                                                         & 12                   & 9                    & 7                    & 9                    & 8                    & 14                   & 12                   & 10                   & 8                    & 7                    &  & 96                              & 9.6                               \\
			\multicolumn{2}{l}{Percentage Mismatch Companies}                                                      & \textbf{46.2}        & \textbf{33.3}        & \textbf{25.9}        & \textbf{36.0}        & \textbf{30.8}        & \textbf{51.9}        & \textbf{44.4}        & \textbf{37.0}        & \textbf{44.4}        & \textbf{38.9}        &  & 38.7                            & {\color{red}\textbf{38.9}}                              \\ \hline
		\end{tabular}
		\label{tab:test1_cdp_scope1_oilgas}
	}
\end{minipage}
\end{center}
\end{sidewaystable}

\begin{sidewaystable}
\sidewaystablefn
\begin{center}
\begin{minipage}{\textheight}
	{\footnotesize
		\caption{Test 1 results on Oil \& Gas companies. N/A-Not Available. FF-Filter Failed}
\begin{tabular}{clcccccccc}
	\hline
	\textbf{}   & \multicolumn{1}{c}{\multirow{2}{*}{\textbf{Organisation}}} & \multirow{2}{*}{\textbf{Country}} & \multicolumn{5}{c}{\textbf{Breakdown}}                                                     & \multirow{2}{*}{\textbf{Total}} & \multirow{2}{*}{\textbf{Percentage}}  \\ \cline{4-8}
	\textbf{}   & \multicolumn{1}{c}{}                                       &                                   & \textbf{Activity} & \textbf{Business} & \textbf{Facility} & \textbf{GHG} & \textbf{Region} &                                 &                                                                         \\ \hline
	\textbf{1}  & Royal Dutch Shell                                          & Netherlands                       & n/a               & 8/9               & n/a               & 10/10        & 9/10            & 27/29                           & 93.1\%                                                      \\
	\textbf{2}  & PTT                                                        & Thailand                          & 3/4               & 4/4               & n/a               & 5/5          & 0/1             & 12/14                           & 85.7\%                                                                  \\
	\textbf{3}  & Woodside Petroleum                                         & Australia                         & 0/1               & n/a               & 2/2               & 3/3          & 3/5             & 8/11                            & 72.7\%                                                                  \\
	\textbf{4}  & Occidental Petroleum Corporation                           & USA                               & n/a               & 3/3               & n/a               & 3/7          & 5/9             & 11/19                           & 57.9\%                                                                  \\
	\textbf{5}  & Cenovus Energy                                             & Canada                            & 3/7               & 1/7               & 1/2               & 7/8          & n/a             & 12/24                           & 50\%                                                                    \\
	\textbf{6}  & Neste Oyj                                                  & Finland                           & 4/7               & 3/8               & 7/8               & 2/7          & 2/8             & 18/38                           & 47.4\%                                                                  \\
	\textbf{7}  & Devon Energy Corporation                                   & USA                               & 5/6               & 1/8               & ff               & 8/8          & 0/8             & 14/30                           & 46.7\%                                                                  \\
	\textbf{8}  & Chevron Corporation                                        & USA                               & 1/3               & 4/8               & n/a               & 2/8          & 2/8             & 9/27                            & 33.3\%                                                                  \\
	\textbf{9}  & Oil \& Natural Gas                                         & India                             & n/a               & 0/2               & n/a               & 2/2          & 0/2             & 2/6                             & 33.3\%                                                                  \\
	\textbf{10} & Marathon Oil Corporation                                   & USA                               & n/a               & 0/2               & n/a               & 2/3          & 0/1             & 2/6                             & 33.3\%                                                                  \\
	\textbf{11} & Ecopetrol                                                  & Colombia                          & 0/4               & 2/6               & 2/5               & 2/5          & n/a             & 6/20                            & 30\%                                                                    \\
	\textbf{12} & Petrobras                                                  & Brazil                            & n/a               & 3/10              & n/a               & 4/10         & 1/10            & 8/30                            & 26.7\%                                                                  \\
	\textbf{13} & Inpex Corporation                                          & Japan                             & n/a               & 1/4               & 1/3               & 1/4          & 2/8             & 5/19                            & 26.3\%                                                                  \\
	\textbf{14} & Novatek                                                    & Russia                            & n/a               & 2/7               & 1/1               & 0/1          & 0/3             & 3/12                            & 25\%                                                                    \\
	\textbf{15} & eni                                                        & Italy                             & 2/9               & 1/10              & 4/6               & 2/10         & 1/10            & 10/45                           & 22.2\%                                                                  \\
	\textbf{16} & Canadian Natural Resources                                 & Canada                            & n/a               & 1/6               & n/a               & 2/3          & 1/10            & 4/19                            & 21.1\%                                                                  \\
	\textbf{17} & Lukoil                                                     & Russia                            & 0/1               & n/a               & n/a               & 0/2          & 1/2             & 1/5                             & 20\%                                                                    \\
	\textbf{18} & BP                                                         & UK                                & n/a               & 2/8               & n/a               & 1/8          & 1/8             & 4/24                            & 16.7\%                                                                  \\
	\textbf{19} & OMV                                                        & Austria                           & 1/1               & 2/10              & 0/1               & 0/10         & 1/10            & 4/32                            & 12.5\%                                                                  \\
	\textbf{20} & Oil Search                                                 & Australia                         & n/a               & n/a               & 0/4               & 1/5          & 0/2             & 1/11                            & 9.1\%                                                                   \\
	\textbf{21} & Repsol                                                     & Spain                             & 0/7               & 1/9               & 1/9               & 1/10         & 1/10            & 4/45                            & 8.9\%                                                                   \\
	\textbf{22} & Hess Corporation                                           & USA                               & n/a               & 0/7               & 0/2               & 0/9          & 2/10            & 2/28                            & 7.1\%                                                                   \\
	\textbf{23} & Suncor Energy                                              & Canada                            & 0/7               & 2/8               & 1/10              & 0/10         & 0/10            & 3/45                            & 6.7\%                                                                   \\
	\textbf{24} & EOG Resources                                              & USA                               & 0/6               & n/a               & n/a               & 1/2          & 0/7             & 1/15                            & 6.7\%                                                                   \\
	\textbf{25} & Imperial Oil                                               & Canada                            & n/a               & 1/7               & n/a               & 0/8          & 0/1             & 1/16                            & 6.3\%                                                                   \\
	\textbf{26} & Exxon Mobil Corporation                                    & USA                               & n/a               & 0/7               & n/a               & 1/7          & 0/8             & 1/22                            & 4.5\%                                                                   \\
	\textbf{27} & ConocoPhillips                                             & USA                               & n/a               & 1/10              & n/a               & 0/10         & 0/10            & 1/30                            & 3.3\%                                                                   \\
	\textbf{28} & Equinor                                                    & Norway                            & n/a               & 0/9               & 1/2               & 0/10         & 0/9             & 1/30                            & 3.3\%                                                                   \\
	\textbf{29} & Galp Energia                                               & Portugal                          & 0/2               & 0/6               & n/a               & 0/8          & 0/8             & 0/24                            & 0\%                                                                     \\
	\textbf{30} & Total                                                      & France                            & n/a               & 0/8               & n/a               & 0/8          & 0/8             & 0/24                            & 0\%                                                                     \\
	\textbf{31} & Apache Corporation                                         & USA                               & 0/3               & n/a               & n/a               & 0/6          & 0/6             & 0/15                            & 0\%                                                                     \\
	\textbf{32} & Noble Energy                                               & USA                               & 0/6               & ff               & n/a               & ff          & 0/6             & 0/12                            & 0\%                                                                     \\
	\textbf{33} & CNOOC                                                      & China                             & n/a               & n/a               & 0/1               & 0/1          & 0/2             & 0/4                             & 0\%                                                                     \\ \\

 \multicolumn{3}{c}{\textbf{Total}} &   19/74    &  43/183    &   21/56     &   60/208      &    32/210     &    175/731     &  \\
 
 \hline
\end{tabular}
		\label{tab:test1_scope1}
	}
 \end{minipage}
\end{center}
\end{sidewaystable}

\newpage

\subsection{Acting in Good Faith Test}
The previous section presented a mathematical evaluation that aimed to determine whether the analyzed companies experienced any challenges in utilizing appropriate software to guarantee the accuracy of their emissions disclosure. Thus, taking into account the findings of the preceding section, the current mathematical test aims to identify the intentions of these companies when such errors were committed (see Methods). In other words, this examination transcends the identification of inconsistencies in their reported emissions and aims to determine whether these firms adhered to the ‘Precautionary Principle' as mandated by the EU's Paris Aligned Benchmarks for their emission breakdown reporting.

The overall performance and companies ranking, similarly to the previous section, are shown in \Cref{tab:test2_cdp_scope1_oilgas,tab:test2_scope1}, respectively. In particular, as seen in \Cref{tab:test2_cdp_scope1_oilgas}, we found GHG emissions to represent downward biased unbalanced internal bookkeeping in $15.5\%$ of the cases, worst in 2015 ($25.9\%$), best in 2013 ($4\%$). This implies that out of the 9.6 companies that exhibited inconsistent internal bookkeeping on average over the 10-year period analyzed in the previous mathematical examination 
(see \Cref{tab:test1_cdp_scope1_oilgas}), roughly 40\% (equivalent to 3.8 companies) failed to adhere to the ‘Precautionary Principle', indicating that errors were made in favor of the company rather than in favor of the planet.

\begin{sidewaystable}
\sidewaystablefn%
\begin{center}
\begin{minipage}{\textheight}
	{\footnotesize
		\caption{Overall Oil \& Gas companies performance (Test 2). Note, all reported values (mismatches, total reports, and total companies) have been obtained using the preprocessing procedure described in the methods section.}
\begin{tabular}{llccccccccccccc}
	\hline
	\multicolumn{1}{c}{\multirow{2}{*}{\textbf{Breakdown}}} & \multicolumn{1}{c}{\multirow{2}{*}{\textbf{Parameter}}} & \multicolumn{10}{c}{\textbf{CDP Report}}                                                                                                                      & \textbf{}            & \multirow{2}{*}{\textbf{Total}} & \multirow{2}{*}{\textbf{Average}} \\ \cline{3-12}
	\multicolumn{1}{c}{}                                    & \multicolumn{1}{c}{}                                    & \textbf{2010} & \textbf{2011} & \textbf{2012} & \textbf{2013} & \textbf{2014} & \textbf{2015} & \textbf{2016} & \textbf{2017} & \textbf{2018} & \textbf{2019} & \multicolumn{1}{l}{} &                                 &                                   \\ \hline
	\multirow{3}{*}{\textbf{Activity}}                      & Mismatch                                                &               & 0             & 1             & 1             & 1             & 2             & 1             & 0             & 1             & 0             &                      & 7                               & 0.8                               \\
	& Percentage Mismatch                              &               & 0.0           & 10.0          & 10.0          & 10.0          & 18.2          & 10.0          & 0.0           & 33.3          & 0.0           &                      & 9.5                             & 10.2                              \\
	& Total Reports                            &               & 6             & 10            & 10            & 10            & 11            & 10            & 11            & 3             & 3             &                      & 74                              & 8.2                               \\
	&                                                         &               &               &               &               &               &               &               &               &               &               &                      &                                 &                                   \\
	\multirow{3}{*}{\textbf{Business}}                      & Mismatch                                                & 1             & 2             & 0             & 0             & 4             & 4             & 0             & 2             & 0             & 1             &                      & 14                              & 1.4                               \\
	& Percentage Mismatch                               & 5.6           & 10.0          & 0.0           & 0.0           & 18.2          & 18.2          & 0.0           & 9.5           & 0.0           & 9.1           &                      & 7.7                             & 7.1                               \\
	& Total Reports                             & 18            & 20            & 19            & 18            & 22            & 22            & 22            & 21            & 10            & 11            &                      & 183                             & 18.3                              \\
	&                                                         &               &               &               &               &               &               &               &               &               &               &                      &                                 &                                   \\
	\multirow{3}{*}{\textbf{Facility}}                      & Mismatch                                                & 1             & 1             & 0             & 0             & 0             & 0             & 0             & 0             & 0             & 1             &                      & 3                               & 0.3                               \\
	& Percentage Mismatch                               & 12.5          & 14.3          & 0.0           & 0.0           & 0.0           & 0.0           & 0.0           & 0.0           & 0.0           & 16.7          &                      & 5.4                             & 4.3                               \\
	& Total Reports                             & 8             & 7             & 4             & 4             & 5             & 5             & 6             & 5             & 6             & 6             &                      & 56                              & 5.6                               \\
	&                                                         &               &               &               &               &               &               &               &               &               &               &                      &                                 &                                   \\
	\multirow{3}{*}{\textbf{GHG}}                           & Mismatch                                                & 2             & 2             & 0             & 0             & 3             & 2             & 3             & 3             & 2             & 1             &                      & 18                              & 1.8                               \\
	& Percentage Mismatch                               & 9.1           & 8.7           & 0.0           & 0.0           & 13.6          & 9.5           & 14.3          & 15.0          & 11.1          & 5.6           &                      & 8.7                             & 8.7                               \\
	& Total Reports                             & 22            & 23            & 22            & 21            & 22            & 21            & 21            & 20            & 18            & 18            &                      & 208                             & 20.8                              \\
	&                                                         &               &               &               &               &               &               &               &               &               &               &                      &                                 &                                   \\
	\multirow{3}{*}{\textbf{Region}}                        & Mismatch                                                & 1             & 1             & 1             & 0             & 3             & 3             & 1             & 2             & 2             & 0             &                      & 14                              & 1.4                               \\
	& Percentage Mismatch                               & 4.8           & 4.5           & 4.3           & 0.0           & 13.6          & 13.6          & 4.8           & 9.1           & 11.1          & 0.0           &                      & 6.7                             & 6.6                               \\
	& Total Reports                             & 21            & 22            & 23            & 21            & 22            & 22            & 21            & 22            & 18            & 18            &                      & 210                             & 21.0                              \\
	&                                                         &               &               &               &               &               &               &               &               &               &               &                      &                                 &                                   \\
	\multicolumn{2}{c}{\textbf{Total Mismatches}}                                                                     & 5             & 6             & 2             & 1             & 11            & 11            & 5             & 7             & 5             & 3             &                      & \textbf{56}                     &                                   \\
	\multicolumn{2}{c}{\textbf{Total Reports}}                                                                        & 69            & 78            & 78            & 74            & 81            & 81            & 80            & 79            & 55            & 56            &                      & \textbf{731}                    &                                   \\
	\multicolumn{2}{c}{\textbf{Percentage Mismatch}}                                                                           & 7.2           & 7.7           & 2.6           & 1.4           & 13.6          & 13.6          & 6.3           & 8.9           & 9.1           & 5.4           &                      & \textbf{7.7}                    &                                   \\
	&                                                         &               &               &               &               &               &               &               &               &               &               &                      &                                 &                                   \\ \hline
	\multicolumn{2}{l}{Total Companies}                                                            & \textbf{26}   & \textbf{27}   & \textbf{27}   & \textbf{25}   & \textbf{26}   & \textbf{27}   & \textbf{27}   & \textbf{27}   & \textbf{18}   & \textbf{18}   &                      & 248                             & 24.8                              \\
	\multicolumn{2}{l}{Mismatch Companies}                                                                         & 5             & 3             & 2             & 1             & 5             & 7             & 4             & 4             & 4             & 3             &                      & 38                              & 3.8                               \\
	\multicolumn{2}{l}{Percentage Mismatch Companies}                                                      & \textbf{19.2} & \textbf{11.1} & \textbf{7.4}  & \textbf{4.0}  & \textbf{19.2} & \textbf{25.9} & \textbf{14.8} & \textbf{14.8} & \textbf{22.2} & \textbf{16.7} &                      & 15.3                            & {\color{red}\textbf{15.5}}                              \\ \hline
\end{tabular}
		\label{tab:test2_cdp_scope1_oilgas}
	}
  \end{minipage}
\end{center}
\end{sidewaystable}

\begin{sidewaystable}
\sidewaystablefn
\begin{center}
\begin{minipage}{\textheight}
	{\footnotesize
	\caption{Test 2 results on Oil \& Gas companies. N/A-Not Available. FF-Filter Failed}
\begin{tabular}{clcccccccc}
	\hline
	\textbf{}   & \multicolumn{1}{c}{\multirow{2}{*}{\textbf{Organisation}}} & \multirow{2}{*}{\textbf{Country}} & \multicolumn{5}{c}{\textbf{Breakdown}}                                                     & \multirow{2}{*}{\textbf{Total}} & \multirow{2}{*}{\textbf{Percentage}}  \\ \cline{4-8}
	\textbf{}   & \multicolumn{1}{c}{}                                       &                                   & \textbf{Activity} & \textbf{Business} & \textbf{Facility} & \textbf{GHG} & \textbf{Region} &                                 &                                                                         \\ \hline
	\textbf{1}  & Royal Dutch Shell                                          & Netherlands                       & n/a               & 4/9               & n/a               & 5/10         & 6/10            & 15/29                           & 51.7\%                                                       \\
	\textbf{2}  & Woodside Petroleum                                         & Australia                         & 0/1               & n/a               & 1/2               & 0/3          & 1/5             & 2/11                            & 18.2\%                                                                 \\
	\textbf{3}  & Marathon Oil Corporation                                   & USA                               & n/a               & 0/2               & n/a               & 1/3          & 0/1             & 1/6                             & 16.7\%                                                                  \\
	\textbf{4}  & Ecopetrol                                                  & Colombia                          & 0/4               & 1/6               & 1/5               & 1/5          & n/a             & 3/20                            & 15\%                                                                    \\
	\textbf{5}  & Chevron Corporation                                        & USA                               & 0/3               & 1/8               & n/a               & 2/8          & 1/8             & 4/27                            & 14.8\%                                                                  \\
	\textbf{6}  & PTT                                                        & Thailand                          & 1/4               & 1/4               & n/a               & 0/5          & 0/1             & 2/14                            & 14.3\%                                                                  \\
	\textbf{7}  & Devon Energy Corporation                                   & USA                               & 2/6               & 0/8               & ff               & 2/8          & 0/8             & 4/30                            & 13.3\%                                                                  \\
	\textbf{8}  & Cenovus Energy                                             & Canada                            & 2/7               & 1/7               & 0/2               & 0/8          & n/a             & 3/24                            & 12.5\%                                                                  \\
	\textbf{9}  & Canadian Natural Resources                                 & Canada                            & n/a               & 1/6               & n/a               & 1/3          & 0/10            & 2/19                            & 10.5\%                                                                  \\
	\textbf{10} & Occidental Petroleum Corporation                           & USA                               & n/a               & 0/3               & n/a               & 1/7          & 1/9             & 2/19                            & 10.5\%                                                                  \\
	\textbf{11} & Eni                                                        & Italy                             & 2/9               & 1/10              & 0/6               & 1/10         & 0/10            & 4/45                            & 8.9\%                                                                   \\
	\textbf{12} & BP                                                         & United Kingdom                    & n/a               & 1/8               & n/a               & 0/8          & 1/8             & 2/24                            & 8.3\%                                                                   \\
	\textbf{13} & Novatek                                                    & Russia                            & n/a               & 1/7               & 0/1               & 0/1          & 0/3             & 1/12                            & 8.3\%                                                                   \\
	\textbf{14} & Neste Oyj                                                  & Finland                           & 0/7               & 1/8               & 0/8               & 1/7          & 1/8             & 3/38                            & 7.9\%                                                                   \\
	\textbf{15} & EOG Resources                                              & USA                               & 0/6               & n/a               & n/a               & 1/2          & 0/7             & 1/15                            & 6.7\%                                                                   \\
	\textbf{16} & Imperial Oil                                               & Canada                            & n/a               & 1/7               & n/a               & 0/8          & 0/1             & 1/16                            & 6.3\%                                                                   \\
	\textbf{17} & Inpex Corporation                                          & Japan                             & n/a               & 0/4               & 0/3               & 0/4          & 1/8             & 1/19                            & 5.3\%                                                                   \\
	\textbf{18} & Exxon Mobil Corporation                                    & USA                               & n/a               & 0/7               & n/a               & 1/7          & 0/8             & 1/22                            & 4.5\%                                                                   \\
	\textbf{19} & Hess Corporation                                           & USA                               & n/a               & 0/7               & 0/2               & 0/9          & 1/10            & 1/28                            & 3.6\%                                                                   \\
	\textbf{20} & Petrobras                                                  & Brazil                            & n/a               & 0/10              & n/a               & 1/10         & 0/10            & 1/30                            & 3.3\%                                                                   \\
	\textbf{21} & OMV                                                        & Austria                           & 0/1               & 0/10              & 0/1               & 0/10         & 1/10            & 1/32                            & 3.1\%                                                                   \\
	\textbf{22} & Repsol                                                     & Spain                             & 0/7               & 0/9               & 1/9               & 0/10         & 0/10            & 1/45                            & 2.2\%                                                                   \\
	\textbf{23} & Suncor Energy                                              & Canada                            & 0/7               & 0/8               & 0/10              & 0/10         & 0/10            & 0/45                            & 0\%                                                                     \\
	\textbf{24} & ConocoPhillips                                             & USA                               & n/a               & 0/10              & n/a               & 0/10         & 0/10            & 0/30                            & 0\%                                                                     \\
	\textbf{25} & Equinor                                                    & Norway                            & n/a               & 0/9               & 0/2               & 0/10         & 0/9             & 0/30                            & 0\%                                                                     \\
	\textbf{26} & Galp Energia                                               & Portugal                          & 0/2               & 0/6               & n/a               & 0/8          & 0/8             & 0/24                            & 0\%                                                                     \\
	\textbf{27} & Total                                                      & France                            & n/a               & 0/8               & n/a               & 0/8          & 0/8             & 0/24                            & 0\%                                                                     \\
	\textbf{28} & Apache Corporation                                         & USA                               & 0/3               & n/a               & n/a               & 0/6          & 0/6             & 0/15                            & 0\%                                                                     \\
	\textbf{29} & Noble Energy                                               & USA                               & 0/6               & ff                & n/a               & ff           & 0/6             & 0/12                            & 0\%                                                                     \\
	\textbf{30} & Oil Search                                                 & Australia                         & n/a               & n/a               & 0/4               & 0/5          & 0/2             & 0/11                            & 0\%                                                                     \\
	\textbf{31} & Oil \& Natural Gas                                         & India                             & n/a               & 0/2               & n/a               & 0/2          & 0/2             & 0/6                             & 0\%                                                                     \\
	\textbf{32} & Lukoil                                                     & Russia                            & 0/1               & n/a               & n/a               & 0/2          & 0/2             & 0/5                             & 0\%                                                                     \\
	\textbf{33} & CNOOC                                                      & China                             & n/a               & n/a               & 0/1               & 0/1          & 0/2             & 0/4                             & 0\%                                                                    \\ \\

 \multicolumn{3}{c}{\textbf{Total}} &   7/74    &  14/183    &   3/56     &   18/208      &    14/210     &    56/731     &  \\
 
 \hline
\end{tabular}
	\label{tab:test2_scope1}
}
\end{minipage}
\end{center}
\end{sidewaystable}

Research on sustainability disclosure argues that companies opt for assurance to increase the credibility of the disclosed information~\cite{simnett2009}. Indeed, experimental evidence shows that investors value sustainability assurance~\cite{reimsbach2018}. This increase in credibility can be attributed to the assumption that assured information is, on average, more accurate. Transferred to our setting of carbon emission data, this means when company indicate that their emissions data is assured, one can reasonably expect more consistent emissions breakdowns.

We report descriptive data on carbon assurance as provided by companies via the CDP questionnaire in \Cref{tab:analysed_assurance}, where lower values are better. \Cref{tab:analysed_correlation} shows low correlation coefficients for the correlations of the \textit{Mismatch Percentage}. Additionally, we created binary variables for each sample year, which are one for any indication of assurance and zero otherwise (see \Cref{tab:matrix_correlation_binary}). Overall, we see no convincing evidence of improved carbon breakdowns for companies which provide (higher quality) assurance. That means, the voluntary assurance of carbon emission is not (yet) an effective tool to improve the quality of emissions breakdowns.

\begin{table}[ht!]
	\centering
	{\footnotesize
		\caption{Quality of assurance-related questions as provided by each company via CDP. \textit{Average Score} -- average quality of answers. \textit{N/A Percentage} -- percentage of unanswered questions.}
\begin{tabular}{lccc}
\hline
\multicolumn{1}{c}{\multirow{2}{*}{\textbf{Organisation}}} & \multirow{2}{*}{\textbf{\begin{tabular}[c]{@{}c@{}}Mismatch \\ Percentage\end{tabular}}} & \multicolumn{2}{c}{\textbf{Assurance}}           \\ \cline{3-4} 
\multicolumn{1}{c}{}                                       &                                                                                          & \textit{Average Score} & \textit{N/A Percentage} \\ \hline
Royal Dutch Shell                                          & 93.1                                                                                     & 2.9                    & 0                       \\
PTT                                                        & 85.7                                                                                     & 3.1                    & 40.9                    \\
Woodside Petroleum                                         & 72.7                                                                                     & 2.4                    & 0                       \\
Occidental Petroleum Corporation                           & 57.9                                                                                     & 3.3                    & 13.6                    \\
Cenovus Energy                                             & 50                                                                                       & 2.6                    & 18.2                    \\
Neste Oyj                                              & 47.4                                                                                     & 1.8                    & 0                       \\
Devon Energy Corporation                                   & 45.2                                                                                     & 2.8                    & 9.1                     \\
Chevron Corporation                                        & 33.3                                                                                     & 2.8                    & 18.2                    \\
Marathon Oil Corporation                                   & 33.3                                                                                     & 3.9                    & 68.2                    \\
Oil \& Natural Gas                                         & 33.3                                                                                     & 5.5                    & 90.9                    \\
Ecopetrol                                               & 30                                                                                       & 6.0                    & 72.7                    \\
Petrobras                                                  & 26.7                                                                                     & 3.1                    & 22.7                    \\
Inpex Corporation                                          & 26.3                                                                                     & 2.6                    & 22.7                    \\
Novatek                                                    & 25                                                                                       & 5.3                    & 63.6                    \\
Eni                                                        & 22.2                                                                                     & 2.4                    & 0                       \\
Canadian Natural Resources                                 & 21.1                                                                                     & 2.6                    & 0                       \\
Lukoil                                                  & 20                                                                                       & 3.0                    & 77.3                    \\
BP                                                         & 16.7                                                                                     & 2.8                    & 18.2                    \\
OMV                                                      & 12.5                                                                                     & 2.4                    & 0                       \\
Oil Search                                                 & 9.1                                                                                      & 3.2                    & 59.1                    \\
Repsol                                                 & 8.9                                                                                      & 2.5                    & 0                       \\
Hess Corporation                                           & 7.1                                                                                      & 2.9                    & 0                       \\
Suncor Energy                                          & 6.7                                                                                      & 2.7                    & 0                       \\
EOG Resources                                          & 6.7                                                                                      & 3.8                    & 72.7                    \\
Imperial Oil                                               & 6.3                                                                                      & 1.8                    & 18.2                    \\
Exxon Mobil Corporation                                    & 4.5                                                                                      & 2.4                    & 22.7                    \\
ConocoPhillips                                             & 3.3                                                                                      & 2.6                    & 0                       \\
Equinor                                                    & 3.3                                                                                      & 2.0                    & 81.8                    \\
Total                                                      & 0                                                                                        & 2.5                    & 0                       \\
Galp Energia                                               & 0                                                                                        & 2.6                    & 18.2                    \\
Apache Corporation                                         & 0                                                                                        & 2.8                    & 45.5                    \\
Noble Energy                                           & 0                                                                                        & 4.9                    & 63.6                    \\
CNOOC                                                      & 0                                                                                        & 3.0                    & 86.4                    \\ \hline
\end{tabular}
		\label{tab:analysed_assurance}
	}
\end{table}

\begin{table}[ht!]
	\centering
	{\footnotesize
		\caption{Correlation between studied variables. \textit{Average Score} -- average quality of answers. \textit{N/A Percentage} -- percentage of unanswered questions.}

\begin{tabular}{lccc}
\hline
                    & Mismatch Percentage & Average Score & N/A Percentage \\ \hline
Mismatch Percentage & 1                            &                        &                         \\
Average Score       & 0.03                         & 1                      &                         \\
N/A Percentage      & -0.19                        & 0.63                   & 1                       \\ \hline
\end{tabular}
		\label{tab:analysed_correlation}
	}
\end{table}

\newpage

\section{Conclusions}\label{sec4}

In our analysis of the \textit{Scope~1} emissions reported by companies from the Oil \& Gas industry and their respective breakdowns, we found a considerably large amount of misreporting. First, on average, we find that 38.9\% of the companies do not add up to the sum of \textit{Scope~1} emissions reported. This proportion of mismatches seems especially high when considering that the reporting of breakdowns is voluntary and that validation of the reported amounts is relatively straightforward. While this first analysis provides insights into basic mistakes of bookkeeping, our second test focusses on the potential intentions of companies and finds that in 15.5\% of the cases, the sum of the breakdowns exceeds the total \textit{Scope~1} emissions reported by the company. It seems that in such cases, companies do not follow the precautionary principle in their reporting of total \textit{Scope~1} emissions.

Our results are especially surprising and noteworthy, because our analysis focusses on voluntary reporting to the CDP, which is perceived as one of the most reliable sources of corporate climate-related information. Other data providers and rating agencies rely on data provided via the CDP and carbon emissions are a central indicator of companies' climate impacts. Misreporting on a central matter, such as \textit{Scope~1} emissions, provides a cautionary tale to users of such data. A certain amount of noise seems unavoidable - at least under the currently existing systems. It should also be noted that \textit{Scope~1} emissions only constitute a small, yet very easiest-to-report fraction of the GHG emissions O\&G companies are responsible for through their activities, adding to the overall concerns surrounding companies' low reporting standards. 

The recent development towards more mandatory reporting regulation around the world (e.g., SEC, ISSB, EFRAG) might lead to improvements in disclosure quality. For example, the EFRAG-proposal of European Sustainability Reporting Standards (ESRS) also proposes breakdowns on countries and operating segments in its Application Guidance (paragraphs 52 \& 53) to ESRS E1. Such breakdowns can be useful to better understand transitory risks of companies due to country- or industry-specific developments. This raises the question whether such disclosure mandates, including the requirement for assurance and the connection to fines for misreporting, can help to improve the quality of reported carbon emissions and their respective breakdowns.

\section{Methods}\label{sec2}

We analyse the GHG emissions reported by O\&G companies on the CDP database (see \Cref{tab:analysed_companies_appendix}), which is a not-for-profit organization that runs a global disclosure system for investors, companies, cities, states and regions to manage their environmental impacts~\cite{depoers2016voluntary}.

\subsection*{Data}\label{subsec_database}
The CDP collects corporate climate-related data every year through a questionnaire and its content is managed with prescribed-format answers. The GHG emissions are required to be provided into three different groups or ‘\textit{Scopes}' in the CDP database: \textit{Scope 1}, which are direct emissions owned or controlled by a company, i.e., stationary combustion, mobile combustion, fugitive emissions, and process emissions; \textit{Scope 2}, which are indirect GHG emissions released in the atmosphere, from the consumption of purchased electricity, steam, heat and cooling; \textit{Scope 3}, representing all indirect emissions (not included in \textit{Scope~2}) that occur in the value chain of the reporting company. In addition, when reporting their \textit{Scope~1} emissions to the CDP, companies are encouraged to voluntarily report their total GHG emissions broken down into Activities, Business Units, Facilities, GHG Types, and Regions (see \Cref{tab:s1breakquestions}).

In this article, we focus on 33 O\&G companies from the CDP database between 2010 and 2019. We use the O\&G classification provided by the Transition Pathway Initiative (TPI), which distributes companies according to the management of their GHG emissions and opportunities related to the low-carbon transition. That is, O\&G companies are classified within the following four levels: \textit{i)} Awareness; \textit{ii)} Building Capacity; \textit{iii)} Interacting into Operational Decision Making; \textit{iv)} Strategic Assessment.

\subsection*{Testing Abilities and Intentions}\label{sub_tests}
We analyzed the (breakdowns of) GHG emissions of 33 O\&G companies as reported to the CDP between 2010 and 2019 based on two tests:

\begin{enumerate}
	\item \textbf{Basic Bookkeeping (Test 1)}, which raises concerns about \textit{\underline{ability}}. That is, if companies had a suitable software or just an accurate Excel spreadsheet for their voluntary GHG breakdown reporting, the following equation should hold:
    \begin{equation}\label{eq:test_1}
		\alpha = \beta = \sum_{i=1}^{n} \beta_{i} 
	\end{equation}
	where $\alpha$ is the reported global \textit{Scope~1} emissions (given in metric tonnes CO2e), while $\beta$ indicates the sum of \textit{Scope~1} emissions over the breakdown in $n$ elements.
	
	\item \textbf{Acting in Good Faith (Test 2)}, which raises concerns about \textit{\underline{intentions}}. Here, if companies struggled with a suitable software or an accurate Excel sheet but followed the \textit{Precautionary Principle} (‘If in doubt, err on the side of the planet not on the side of the company’) as required by the EU’s Paris Aligned Benchmarks for their voluntary GHG breakdown reporting, \Cref{eq:test_2} should hold:
	\begin{equation}\label{eq:test_2}
		\alpha \geq \beta
	\end{equation}
\end{enumerate}

We investigate in Test 1 and Test 2, if \Cref{eq:test_1,eq:test_2}, respectively, hold in the entire CDP database between 2010 and 2019 for all 5 breakdown dimensions (Activities, Business Units, Facilities, GHG Types, and Regions) in \textit{Scope~1} emissions. 
 
\begin{figure}[h!]
	\centering
	\includegraphics[scale=0.4]{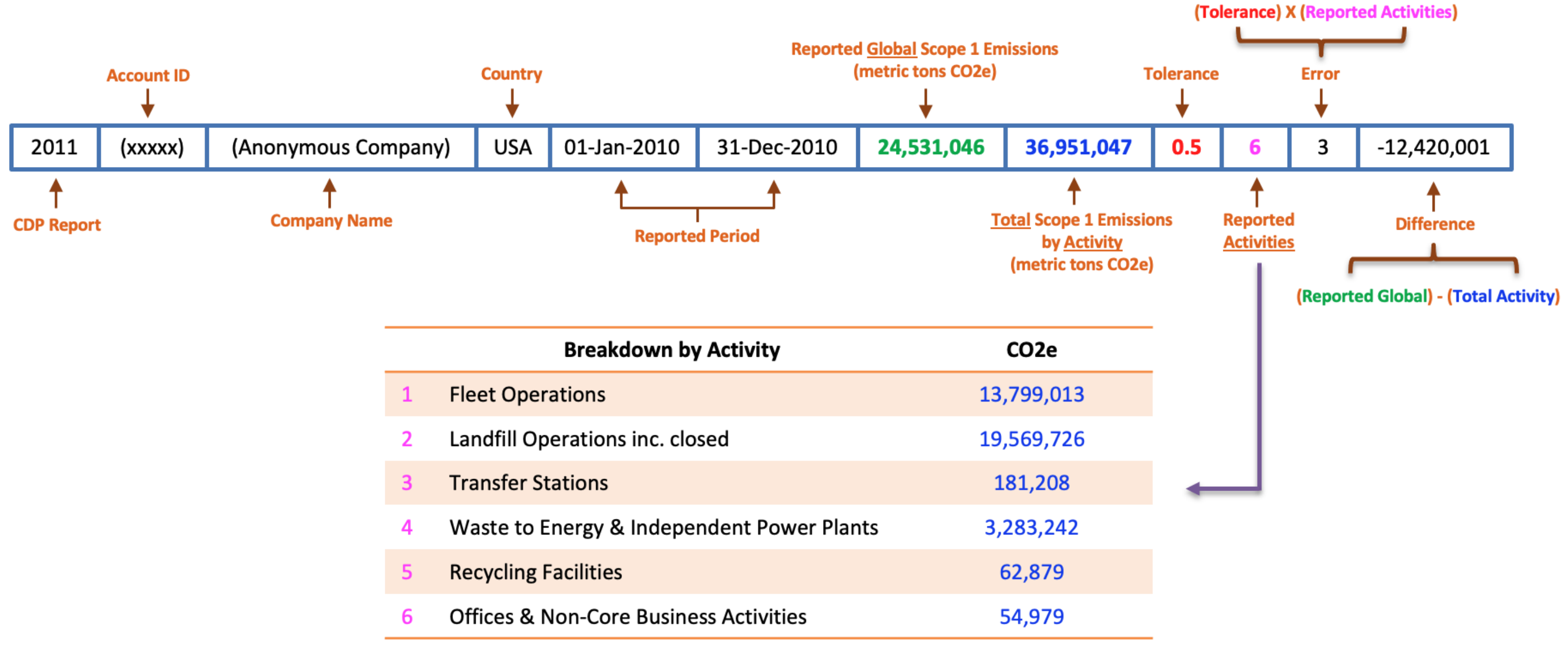}
	\caption{A representative example for testing abilities (Test~1) and intentions (Test~2). The example shows that the analyzed company failed to pass Test~1, as the reported global \textit{Scope~1} emissions (24,531,046 metric tons CO2e) is different from the total sum of the breakdown (36,951,047 metric tons CO2e), i.e., $\alpha\neq\beta$. Similarly, the analyzed company failed to pass Test~2, as the total sum of the breakdown (36,951,047 metric tons CO2e) is greater than the reported global emissions (24,531,046 metric tons CO2e), i.e., $\alpha<\beta$.}
	\label{fig:methodology}
\end{figure}

To illustrate, \Cref{fig:methodology} shows a representative example of the methodology used to validate the proposed tests. Note: \textit{i)} The reporting period is not always equal to the calendar year (e.g., when the fiscal year does not equal the calendar year, and the company chose to report emissions for the fiscal year). We always considered reporting periods of a duration of 12 months. This 12 months period lies within the range from January 1st of the previous CDP year to December 31st of the current CDP year. \textit{ii)} Only numeric values were considered when computing \textit{Total Scope 1 Emissions}, i.e., values like ‘null' or ‘n/a' were treated as 0 and therefore do not increase the total sum on each breakdown. \textit{iii)} Tolerance was computed based on the number of decimals a company used to report its \textit{Scope~1} emissions. For example, some companies reported \textit{Scope~1} emissions with three decimals while others rounded to nearest 1,000 or even 1,000,000 metric tons of \textit{Scope~1} emissions. We considered the number of decimal places on reported global \textit{Scope~1} emissions or rounding to the nearest multiple of 10. In particular, it was assumed that the following companies used the latter method: BP, Exxon Mobil Corporation, Marathon Oil Corporation, Occidental Petroleum Corporation, Oil \& Natural Gas, and Total. 

\bibliography{Main_NatureEnergyAnalysis}
\newpage

\newpage
\section{Supplementary Information}

\begin{table}[ht!]
	\centering
	{\footnotesize
		\caption{Analysed Oil \& Gas companies. \textit{TPI}--Transition Pathway Initiative. \textit{ID}--Carbon Disclosure Project ID, which is a unique number assigned to each company (also known ‘Account') in the CDP database.}\label{tab:analysed_companies_appendix}
		\begin{tabular}{cclcc}
			\hline
			\textbf{}   & \multicolumn{1}{c}{\textbf{ID}} & \multicolumn{1}{c}{\textbf{Company}} & \textbf{Country} & \textbf{TPI Level}  \\ \hline
			\textbf{1}  &  13823 & Oil \& Natural Gas                   & India            & 1                   \\
			& &                                      &                  &                     \\
			\textbf{2}  & 11347 & Marathon Oil Corporation             & USA              & \multirow{6}{*}{2}  \\
			\textbf{3}  & 11043 & Lukoil                              & Russia           &                     \\
			\textbf{4}  & 13838 & Oil Search                           & Australia        &                     \\
			\textbf{5}  & 5767 & EOG Resources                        & USA              &                     \\
			\textbf{6}  & 3527 & CNOOC                                & China            &                     \\
			\textbf{7}  & 13395 & Noble Energy                      & USA              &                     \\
			& &                                      &                  &                     \\
			\textbf{8}  & 15297 & PTT                                  & Thailand         & \multirow{13}{*}{3} \\
			\textbf{9}  & 20771 & Woodside Petroleum                   & Australia        &                     \\
			\textbf{10} & 29789 & Cenovus Energy                    & Canada           &                     \\
			\textbf{11} & 12937 & Neste Oyj                        & Finland          &                     \\
			\textbf{12} & 4678 & Devon Energy Corporation             & USA              &                     \\
			\textbf{13} & 3191 & Chevron Corporation                  & USA              &                     \\
			\textbf{14} & 22341 & Ecopetrol                         & Colombia         &                     \\
			\textbf{15} & 13542 & Novatek                              & Russia           &                     \\
			\textbf{16} & 2667 & Canadian Natural Resources   & Canada           &                     \\
			\textbf{17} & 17929 & Suncor Energy                    & Canada           &                     \\
			\textbf{18} & 8886 & Imperial Oil                         & Canada           &                     \\
			\textbf{19} & 6136 & Exxon Mobil Corporation              & USA              &                     \\
			\textbf{20} & 804 & Apache Corporation                   & USA              &                     \\
			&                                      &                  &                     \\
			\textbf{21} & 16012 & Royal Dutch Shell                    & Netherlands      & \multirow{13}{*}{4} \\
			\textbf{22} & 13649 & Occidental Petroleum Corporation     & USA              &                     \\
			\textbf{23} & 14654 & Petrobras                            & Brazil           &                     \\
			\textbf{24} & 9134 & Inpex Corporation                    & Japan            &                     \\
			\textbf{25} & 2083 & BP                                   & UK               &                     \\
			\textbf{26} & 5634 & Eni                                  & Italy            &                     \\
			\textbf{27} & 13870 & OMV                               & Austria          &                     \\
			\textbf{28} & 15669 & Repsol                           & Spain            &                     \\
			\textbf{29} & 8274 & Hess Corporation                     & USA              &                     \\
			\textbf{30} & 19257 & Total                                & France           &                     \\
			\textbf{31} & 3751 & ConocoPhillips                       & USA              &                     \\
			\textbf{32} & 23132 & Equinor                              & Norway           &                     \\
			\textbf{33} & 7042 & Galp Energia                      & Portugal         &                     \\ \hline
		\end{tabular}
	}
\end{table}

\begin{sidewaystable}
\sidewaystablefn
\begin{center}
\begin{minipage}{\textheight}
\caption{Scope 1 Breakdown questions in CDP database.}\label{tab:s1breakquestions}
	{\tiny
\begin{tabular}{clcl}
\hline
\textbf{CDP Questionnaire}     & \multicolumn{1}{c}{\textbf{Breakdown}} & \textbf{Question}    & \multicolumn{1}{c}{\textbf{Description}}                                                                                                                                                     \\ \hline
\multirow{5}{*}{\textbf{2010}} & Activity                               & N/A                  & N/A                                                                                                                                                                                          \\
                               & Business Division                      & 12.4.                & Please also break down your total gross global Scope 1 emissions by business division                                      \\
                               & Facility                               & 12.5.                & Please also break down your total gross global Scope 1 emissions by facility                                               \\
                               & GHG Type                               & 12.6C3.              & Scope 1 Emissions (Metric tonnes CO2-e)                                                                                                                                                      \\
                               & Region                                 & 12.2.                & Please break down your total gross global Scope 1 emissions by country/region                                                                                         \\
\multicolumn{1}{l}{}           &                                        & \multicolumn{1}{l}{} &                                                                                                                                                                                              \\
\multirow{5}{*}{\textbf{2011}} & Activity                               & 9.2d.                & Please break down your total gross global Scope 1 emissions by activity                                                                                                                      \\
                               & Business Divison                       & 9.2a.                & Please break down your total gross global Scope 1 emissions by business division                                                                                                             \\
                               & Facility                               & 9.2b.                & Please break down your total gross global Scope 1 emissions by facility                                                                                                                      \\
                               & GHG Type                               & 9.2c.                & Please break down your total gross global Scope 1 emissions by GHG type                                                                                                                      \\
                               & Region                                 & 9.1a.                & Please break down your total gross global Scope 1 emissions by country/region                                                                                                                \\
\multicolumn{1}{l}{}           &                                        & \multicolumn{1}{l}{} &                                                                                                                                                                                              \\
\multirow{5}{*}{2012}          & Activity                               & 9.2d.                & Please break down your total gross global Scope 1 emissions by activity                                                                                                                      \\
                               & Business Division                      & 9.2a.                & Please break down your total gross global Scope 1 emissions by business division                                                                                                             \\
                               & Facility                               & 9.2b.                & Please break down your total gross global Scope 1 emissions by facility                                                                                                                      \\
                               & GHG Type                               & 9.2c.                & Please break down your total gross global Scope 1 emissions by GHG type                                                                                                                      \\
                               & Region                                 & 9.1a.                & Please break down your total gross global Scope 1 emissions by country/region                                                                                                                \\
\multicolumn{1}{l}{}           &                                        & \multicolumn{1}{l}{} &                                                                                                                                                                                              \\
\multirow{5}{*}{\textbf{2013}} & Activity                               & 9.2d.                & Please break down your total gross global Scope 1 emissions by activity                                                                                                                      \\
                               & Business Division                      & 9.2a.                & Please break down your total gross global Scope 1 emissions by business division                                                                                                             \\
                               & Facility                               & 9.2b.                & Please break down your total gross global Scope 1 emissions by facility                                                                                                                      \\
                               & GHG Type                               & 9.2c.                & Please break down your total gross global Scope 1 emissions by GHG type                                                                                                                      \\
                               & Region                                 & 9.1a.                & Please break down your total gross global Scope 1 emissions by country/region                                                                                                                \\
\multicolumn{1}{l}{}           &                                        & \multicolumn{1}{l}{} &                                                                                                                                                                                              \\
\multirow{5}{*}{\textbf{2014}} & Activity                               & CC9.2d.              & Please break down your total gross global Scope 1 emissions by activity                                                                                                                      \\
                               & Business Division                      & CC9.2a.              & Please break down your total gross global Scope 1 emissions by business division                                                                                                             \\
                               & Facility                               & CC9.2b.              & Please break down your total gross global Scope 1 emissions by facility                                                                                                                      \\
                               & GHG Type                               & CC9.2c.              & Please break down your total gross global Scope 1 emissions by GHG type                                                                                                                      \\
                               & Region                                 & CC9.1a.              & Please break down your total gross global Scope 1 emissions by country/region                                                                                                                \\
\multicolumn{1}{l}{}           &                                        & \multicolumn{1}{l}{} &                                                                                                                                                                                              \\
\multirow{5}{*}{\textbf{2015}} & Activity                               & CC9.2d.              & Please break down your total gross global Scope 1 emissions by activity                                                                                                                      \\
                               & Business Division                      & CC9.2a.              & Please break down your total gross global Scope 1 emissions by business division                                                                                                             \\
                               & Facility                               & CC9.2b.              & Please break down your total gross global Scope 1 emissions by facility                                                                                                                      \\
                               & GHG Type                               & CC9.2c.              & Please break down your total gross global Scope 1 emissions by GHG type                                                                                                                      \\
                               & Region                                 & CC9.1a.              & Please break down your total gross global Scope 1 emissions by country/region                                                                                                                \\
\multicolumn{1}{l}{}           &                                        & \multicolumn{1}{l}{} &                                                                                                                                                                                              \\
\multirow{5}{*}{\textbf{2016}} & Activity                               & CC9.2d.              & Please break down your total gross global Scope 1 emissions by activity                                                                                                                      \\
                               & Business Division                      & CC9.2a.              & Please break down your total gross global Scope 1 emissions by business division                                                                                                             \\
                               & Facility                               & CC9.2b.              & Please break down your total gross global Scope 1 emissions by facility                                                                                                                      \\
                               & GHG Type                               & CC9.2c.              & Please break down your total gross global Scope 1 emissions by GHG type                                                                                                                      \\
                               & Region                                 & CC9.1a.              & Please break down your total gross global Scope 1 emissions by country/region                                                                                                                \\
\multicolumn{1}{l}{}           &                                        & \multicolumn{1}{l}{} &                                                                                                                                                                                              \\
\multirow{5}{*}{\textbf{2017}} & Activity                               & CC9.2d C2            & Please break down your total gross global Scope 1 emissions by activity                                                                             \\
                               & Business Division                      & CC9.2a C2            & Please break down your total gross global Scope 1 emissions by business division                                                                    \\
                               & Facility                               & CC9.2b C2            & Please break down your total gross global Scope 1 emissions by facility                                                                             \\
                               & GHG Type                               & CC9.2c C2            & Please break down your total gross global Scope 1 emissions by GHG type                                                                             \\
                               & Region                                 & CC9.1a C2            & Please break down your total gross global Scope 1 emissions by country/region                                                                                   \\
\multicolumn{1}{l}{}           &                                        & \multicolumn{1}{l}{} &                                                                                                                                                                                              \\
\multirow{5}{*}{\textbf{2018}} & Activity                               & C7.3c\_C2            & Break down your total gross global Scope 1 emissions by business activity                                                                             \\
                               & Business Division                      & C7.3a\_C2            & Break down your total gross global Scope 1 emissions by business division                                                                              \\
                               & Facility                               & C7.3b\_C2            & Break down your total gross global Scope 1 emissions by business facility                                                                             \\
                               & GHG Type                               & C7.1a\_C2            & Break down your total gross global Scope 1 emissions by greenhouse gas type \\
                               & Region                                 & C7.2\_C2             & Break down your total gross global Scope 1 emissions by country/region                                                                                \\
\multicolumn{1}{l}{}           &                                        & \multicolumn{1}{l}{} &                                                                                                                                                                                              \\
\multirow{5}{*}{\textbf{2019}} & Activity                               & C7.3c\_C2            & Break down your total gross global Scope 1 emissions by business activity                                                                             \\
                               & Business Division                      & C7.3a\_C2            & Break down your total gross global Scope 1 emissions by business division                                                                              \\
                               & Facility                               & C7.3b\_C2            & Break down your total gross global Scope 1 emissions by business facility                                                                             \\
                               & GHG Type                               & C7.1a\_C2            & Break down your total gross global Scope 1 emissions by greenhouse gas type \\
                               & Region                                 & C7.2\_C2             & Break down your total gross global Scope 1 emissions by country/region                                                                                \\ \hline
\end{tabular}
	}
\end{minipage}
\end{center}
\end{sidewaystable}

\begin{sidewaystable}
\sidewaystablefn
\begin{center}
\begin{minipage}{\textheight}
	{\footnotesize
		\caption{Correlation Matrix (Binary Score). AS-Assurance Score. MP-Mismatch Percentage.}
\begin{tabular}{lcccccccccc}
\hline
                   & \textbf{AS 2010} & \textbf{AS 2011-I} & \textbf{AS 2012-I} & \textbf{AS 2013-I} & \textbf{AS 2014-I} & \textbf{AS 2015-I} & \textbf{AS 2016-I} & \textbf{AS 2017-I} & \textbf{AS 2018-I} & \textbf{AS 2019-I} \\ \hline
\textbf{AS 2010}   & 1                &                    &                    &                    &                    &                    &                    &                    &                    &                    \\
\textbf{MP 2010}   & 0.1              &                    &                    &                    &                    &                    &                    &                    &                    &                    \\
\textbf{AS 2011-I} & 0.82             & 1                  &                    &                    &                    &                    &                    &                    &                    &                    \\
\textbf{MP 2011}   & -0.15            & -0.12              &                    &                    &                    &                    &                    &                    &                    &                    \\
\textbf{AS 2012-I} & 0.69             & 0.77               & 1                  &                    &                    &                    &                    &                    &                    &                    \\
\textbf{MP 2012}   & 0.14             & 0.16               & 0.16               &                    &                    &                    &                    &                    &                    &                    \\
\textbf{AS 2013-I} & 0.43             & 0.52               & 0.73               & 1                  &                    &                    &                    &                    &                    &                    \\
\textbf{MP 2013}   & 0.03             & 0.06               & 0.09               & 0.37               &                    &                    &                    &                    &                    &                    \\
\textbf{AS 2014-I} & 0.56             & 0.65               & 0.86               & 0.86               & 1                  &                    &                    &                    &                    &                    \\
\textbf{MP 2014}   & 0.12             & 0.15               & 0.18               & 0.18               & 0.18               &                    &                    &                    &                    &                    \\
\textbf{AS 2015-I} & 0.43             & 0.52               & 0.73               & 0.86               & 0.86               & 1                  &                    &                    &                    &                    \\
\textbf{MP 2015}   & 0.17             & 0.21               & 0.19               & 0.51               & 0.38               & 0.32               &                    &                    &                    &                    \\
\textbf{AS 2016-I} & 0.37             & 0.58               & 0.53               & 0.67               & 0.67               & 0.8                & 1                  &                    &                    &                    \\
\textbf{MP 2016}   & 0.08             & 0.12               & 0.12               & 0.25               & 0.12               & 0.31               & 0.44               &                    &                    &                    \\
\textbf{AS 2017-I} & 0.3              & 0.52               & 0.59               & 0.73               & 0.73               & 0.86               & 0.94               & 1                  &                    &                    \\
\textbf{MP 2017}   & 0.19             & 0.28               & 0.2                & 0.39               & 0.2                & 0.2                & 0.23               & 0.2                &                    &                    \\
\textbf{AS 2018-I} & 0.06             & 0.14               & 0.22               & 0.35               & 0.35               & 0.35               & 0.42               & 0.48               & 1                  &                    \\
\textbf{MP 2018}   & 0.12             & 0.15               & 0.1                & 0.25               & 0.1                & 0.05               & 0.08               & 0.05               & 0.4                &                    \\
\textbf{AS 2019-I} & -0.17            & -0.07              & -0.05              & 0.09               & 0.09               & 0.09               & 0.19               & 0.23               & 0.82               & 1                  \\
\textbf{MP 2019}   & 0.26             & 0.28               & 0.25               & 0.25               & 0.25               & 0.04               & 0.07               & 0.04               & 0.3                & 0.3                \\ \hline
\end{tabular}
\label{tab:matrix_correlation_binary} }
\end{minipage}
\end{center}
\end{sidewaystable}

\end{document}